\documentclass[twocolumn,journal]{IEEEtran}
\usepackage{pifont}
\usepackage{stmaryrd}
\usepackage{amsmath}
\usepackage{amsfonts}
\usepackage{amssymb}
\usepackage{bbm}
\usepackage{mathrsfs}
\usepackage{tipa}
\usepackage{cases}
\usepackage{mathbbol}
\usepackage[dvips]{graphicx}
\usepackage{algorithm}
\usepackage{algorithmic}
\usepackage{caption}
\usepackage{amsfonts}
\usepackage{amssymb}
\usepackage{txfonts}
\usepackage{mathrsfs}
\usepackage[dvips]{graphicx}
\usepackage{verbatim}
\usepackage{subfigure}
\usepackage{balance}
\usepackage{booktabs}
\usepackage{fancybox}
\usepackage{bm}
\usepackage{extarrows}
\usepackage{algorithm}
\usepackage{algorithmic}
\usepackage{multirow}
\usepackage{array}
\usepackage{graphicx}
\usepackage{epstopdf}
\usepackage{footmisc}
\usepackage{caption}
\usepackage{mathtools}
\usepackage{comment}
\usepackage{color}
\usepackage{dsfont}
\allowdisplaybreaks[4]
\newtheorem{theorem}{Theorem}
\newtheorem{lemma}{Lemma}

\begin{document}

\title{\LARGE Secure UAV Random Networks With Minimum Safety Distance}

\author{Jiawei Lyu and Hui-Ming Wang,~\IEEEmembership{Senior Member,~IEEE}
\thanks{

\copyright~2015 IEEE. Personal use of this material is permitted. Permission from IEEE must be obtained for all other uses, in any current or future media, including reprinting/republishing this material for advertising or promotional purposes, creating new collective works, for resale or redistribution to servers or lists, or reuse of any copyrighted component of this work in other works. 

The authors are with the School of Information and Communications Engineering, and also with the Ministry of Education Key Laboratory for Intelligent Networks and Network Security, Xi'an Jiaotong University, Xi'an, 710049, China (email: xjtu-Lyujw@outlook.com, xjbswhm@gmail.com). 

Digital Object Identifier 10.1109/TVT.2021.3057786
}
}
\maketitle

\begin{abstract}
In this correspondence, we study the physical layer security in a stochastic unmanned aerial vehicles (UAVs) network from a network-wide perspective, where the locations of UAVs are modeled as a Mat$\acute{\text{e}}$rn hard-core point process (MHCPP) to characterize the minimum safety distance between UAVs, and the locations of users and eavesdroppers are modeled as a Poisson cluster process and a Poisson point process, respectively. UAVs adopt zero-forcing precoding to serve multiple ground users and emit artificial noise to combat eavesdropping. We derive the approximations for the coverage probability and secrecy probability of a typical user, with which we derive the secrecy throughput of the whole network. Numerical results show the analytical results can well approximate the simulation results. Impacts of parameters on the secrecy performance are shown. 
\end{abstract}


\section{Introduction}\label{Introduc}

Unmanned aerial vehicles (UAVs) have been considered to play an important role in future communication systems. It is promising to apply UAVs into a wide range of communication scenarios due to their characteristics of flexible deployment, low acquisition and maintenance costs, high maneuverability and hovering ability \cite{UAVWirlesPLSecCommunOverviewResDirec}, e.g., UAVs acting as relays to enable blocked communications \cite{NovDistrAlgoForPhaSynInUAV} and collecting datas from ground Internet of Things devices \cite {QuaOptUplPowerEnaGreenURLLCMobilUAVIoTNetPerturBasedAppro}.
However, UAV communication networks are vulnerable to eavesdropping due to the broadcast nature of wireless channels and especially the line-of-sight (LoS) propagations \cite{UAVWirlesPLSecCommunOverviewResDirec}. The high mobility characteristic of UAVs makes the key management and distribution in traditional cryptography-based schemes more difficult. To tackle this challenge, physical layer security (PLS) exploiting the wireless channel characteristics to secure communications has great potentials to improve the security of UAVs networks. 

There are many endeavors to achieve PLS in UAVs networks \cite{UAVWirlesPLSecCommunOverviewResDirec}, e.g., by employing UAVs as mobile relays \cite{ImprvPLSUAVMobilRelay} and friendly jammers \cite{ImprvPLSUAVFriJamUnknEavesLoc}, and by using joint power allocation and trajectory design \cite{UAVSecDonlnkNOMATransASecUserOriPers}. However, these works mostly consider only point-to-point communication scenarios and do not study the secrecy performance from a perspective of the whole UAVs network. UAVs should be organized to a network to provide services in many applications. Therefore, it is improtant to study the PLS issue from a network-wide perspective. 
To characterize the location distribution of UAVs in a large-scale UAV network and study the average network-wide performance, stochastic geometry is a useful tool \cite{StoGeoWirlesNetVol1Theo}, and have been widely used in many works, e.g., \cite{SecTransLargeScaUAVWirlesNet}-\cite{SecRateAnaUAVmmWavNetUsinMHCPP}. Yao \emph{et al}. \cite{SecTransLargeScaUAVWirlesNet} studied the secrecy performance in a UAV network modeled as a Poisson point process (PPP) with security guard zone technique. Kim \emph{et al}. \cite{MulLayerUAVNetModelPerformAnaly} studied the successful transmission probability and area spectral efficiency (ASE) of a multi-layer aerial networks with UAVs in each layer modeled as PPPs. Yi \emph{et al}. \cite{ClusteredUAVNetmmWaveCommunStoGeoView} evaluated the coverage probability and ASE in a three-dimensional UAV network at mmWave bands where UAVs and users are distributed according to a Poisson cluster process (PCP). However, PPP and PCP models adopted in all these works do not possess the constraint of a minimum safety distance between two UAVs to avoid collisions among UAVs. In other words, the distance between two UAVs can be arbitrarily small. Considering the safety of the UAV deployment, it is obviously impractical.

In \cite{SecRateAnaUAVmmWavNetUsinMHCPP}, UAV base stations (BSs) distributed as a Mat$\acute{\text{e}}$rn hard-core point process (MHCPP) have been studied.
The MHCPP is a repulsive point process possessing a minimum distance between any two nodes,  which makes it a more realistic model to characterize the practical deployment of a UAV network. The authors in  \cite{SecRateAnaUAVmmWavNetUsinMHCPP} analyzed the rate of a typical user at the mmWave band, where it assumes each UAV only serves one user. This assumption is not very realistic for a UAV aerial BS. Furthermore, it proposes to randomly use part of UAVs to transmit jamming with all their power. Since jamming UAVs are randomly distributed, they are possible to be closer to the typical user than the serving UAV is, which will significantly deteriorate the coverage performance. There also lacks the analysis for a network-wide secrecy performance metric. This motivates us to consider the secrecy throughput of a UAV network modeled as an MHCPP with each UAV serving multiple users under some more efficient PLS scheme.

In this correspondence, we study the PLS in a UAV network where each UAV adopts zero-forcing (ZF) precoding and artificial noise-aided (AN-aided) transmission to serve multiple clustered users and meanwhile confound randomly distributed eavesdroppers (Eves). We model the locations of UAVs as an MHCPP, and the clustering feature of ground users is characterized by a PCP. The main contributions of this work are: 1) we analyze the probability of a node belonging to the MHCPP, with which the MHCPP can be approximated by a PPP to make the analysis tractable. 2) Employing this probability, we derive the approximations for the coverage probability (CP) and the secrecy probability (SP) of the typical user, and the network-wide secrecy throughput (ST) under this novel network model. 3) We validate the approximation in the simulation and develop some practical insights into the network design.

\begin{figure}[!t]
\begin{center}
\centering
\includegraphics[width=2.9 in]{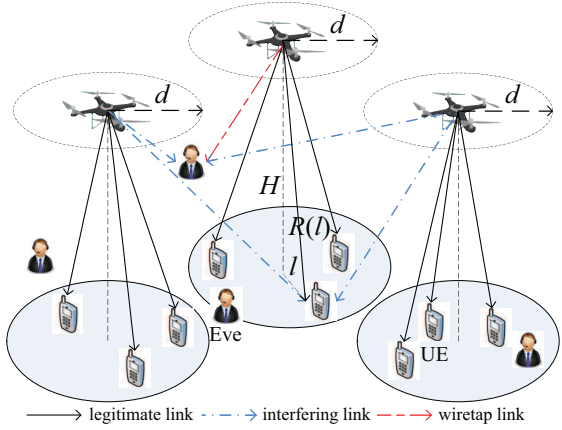}
\end{center}
\caption{Illustration of the considered UAV network.}
\label{SystemModel}
\end{figure}

\section{System Model}\label{system-model}
We consider the security issue in a UAV network where each UAV has several clustered ground users to serve simultaneously as an aerial BS, while there are also randomly located Eves attempting to intercept information from UAVs. An illustration of a network snapshot is depicted in Fig.~\ref{SystemModel}.

\subsection{Spatial Distribution Model}
\label{spatialdistributionmodel}
We assume all UAVs are elevated at the same altitude $H$ and distributed according to an MHCPP, denoted by $\Phi_U$, where the minimum distance between any two UAVs is $d$. The locations of Eves on the ground are modeled by an independent homogeneous PPP $\Phi_E$ with intensity $\lambda_e$. Users are assumed to be distributed according to a symmetric normal distribution around the projections on the ground of their serving UAVs with variance $\sigma^2$. The PDF of the location $x_u$ of a user served by the UAV $u$ is
\begin{equation}
\label{userdistribution}
{f}\left( {x_u}\middle\vert{u} \right) = \frac{1}{2\pi {\sigma^2}}\exp \left({ - \frac{{{\left\lVert{{x_u} - {u}} \right\rVert}^2}}{{2{\sigma^2}}}} \right).
\end{equation}

\subsection{Channel Model}
\label{channelmodel}
Wireless channels are assumed to undergo a small-scale Rayleigh fading and a large-scale path loss. Each UAV possesses $M$ antennas to serve $N$ users ${\left(1\leq N\leq M-1\right)}$ simultaneously while each user and Eve have a single antenna. The small-scale fading vectors from a UAV $u_i$ to its $j$-th user $x_{u_i,j}$ and an Eve $e$ which eavesdrops $u_i$ are denoted by $\mathbf{h}_{i,j}\in\mathbb{C}^{M\times1}$ and $\mathbf{h}_{i,e}\in\mathbb{C}^{M\times1},$ respectively, while the small-scale fading vectors from an interfering UAV $u_i$ to a user $x_{u_k,j}$ ${\left(k\neq i\right)}$ and an Eve $e$ are denoted by $\mathbf{g}_{i,j}\in\mathbb{C}^{M\times1}$ and $\mathbf{g}_{i,e}\in\mathbb{C}^{M\times1},$ respectively. $\mathbf{h}_{i,j}$, $\mathbf{h}_{i,e}$, $\mathbf{g}_{i,j}$ and $\mathbf{g}_{i,e}$ have independent and identically distributed (i.i.d.) entries obeying $\mathcal{CN}\left(0,1\right).$ 
For the large-scale path loss, we adopt the hybrid model proposed in \cite{ImprvPLSUAVFriJamUnknEavesLoc}, \cite{OptLAPAltiMaxCov}, \cite{SecCovrtCommunAgaUAVSurViaMulHopNet}. Specifically, each link from a UAV to a user or an Eve can be LoS with the probability $\mathcal{P}_L$ and be non-line-of-sight (NLoS) with the probability $\mathcal{P}_N=1-\mathcal{P}_L$. The probability $\mathcal{P}_L$ depends on the propagation environment and the elevation angle, which is expressed as \cite{ImprvPLSUAVFriJamUnknEavesLoc}, \cite{OptLAPAltiMaxCov}, \cite{SecCovrtCommunAgaUAVSurViaMulHopNet}  
\begin{equation}
\label{LoSprob}
\mathcal{P}_L{\left(l_i\right)}=\frac{1}{1+a\exp{\left[-b{\left(\frac{180}{\pi}arctan{\left(\frac{H}{l_i}\right)}-a\right)}\right]}},
\end{equation}
where $l_i$ is the horizontal distance from the UAV $u_i$ to the receiver, $a$ and $b$ are environment dependent constants. LoS links and NLoS links possess different path loss coefficients $L_p$, which can be expressed as \cite{ImprvPLSUAVFriJamUnknEavesLoc}, \cite{OptLAPAltiMaxCov}, \cite{SecCovrtCommunAgaUAVSurViaMulHopNet} 
\begin{equation}
L_p{\left(l_i\right)} = \begin{dcases}
\eta_L\xi R{\left(l_i\right)}^{-\alpha_L}, & \text{LoS links with probability $\mathcal{P}_L{\left(l_i\right)}$} \\
\eta_N\xi R{\left(l_i\right)}^{-\alpha_N}, & \text{NLoS links with probability $\mathcal{P}_N{\left(l_i\right)}$},
\end{dcases}
\end{equation}
where $\eta_L$ and $\eta_N$ denote the excessive path loss coefficients for LoS and NLoS links, respectively, $\xi$ is the path loss at the reference distance $1$m, $\alpha_L$ and $\alpha_N$ denote the path loss exponents for LoS and NLoS links, respectively, and $R{\left(l_i\right)}=\sqrt{H^2+l_i^2}$ is the link distance. Besides, the thermal noise power at users and Eves are denoted by $\sigma_x^2$ and $\sigma_e^2$, respectively.

\subsection{Artificial-noise-aided Zero-forcing Transmission}
\label{ANZFTransmission} 
UAVs implement ZF precoding to serve multiple active users in the same cluster \footnote{ZF precoding is proven to be able to achieve asymptotically optimal throughput in the multi-user multiple-input multiple-output system \cite{OnOptMulanteBroadScheUsinZFBeam}, \cite{ZFMethForDwnSpaMultiplexMultiusrMIMOChan}. The computational complexity of ZF precoding is acceptable since UAVs do not have a large amount of antennas.}. Meanwhile, the AN-aided transmission strategy is adopted to confuse Eves \cite{PhyLaySecInHeteCellNet}. The signal vector from the UAV $u_i$ is 
\begin{equation}
\label{signal}
\mathbf{y}_i=\sqrt{\phi P}\mathbf{W}_i\mathbf{s}_i+\sqrt{{\left(1-\phi\right)}P}\mathbf{G}_i\mathbf{v}_i,
\end{equation}
where $\mathbf{s}_i\in\mathbb{C}^{N\times1}$ and $\mathbf{v}_i\in\mathbb{C}^{\left(M-N\right)\times1}$ are the information-bearing signal and an AN vector with i.i.d. entries obeying $\mathcal{CN}\left(0,\frac{1}{N}\right)$ and $\mathcal{CN}\left( {0,\frac{1}{{M-N}}} \right)$, respectively, and $\phi\in\left({0,1}\right)$ is the ratio of the information signal power to the total transmission power $P$. $\mathbf{W}_i\in\mathbb{C}^{M\times N}$ is the ZF precoding matrix consisting of the normalized column vectors of the pseudo-inverse matrix $\overline{\mathbf{H}}_i^{\dag}{\left(\overline{\mathbf{H}}_i\overline{\mathbf{H}}_i^{\dag}\right)}^{-1}\in\mathbb{C}^{M\times N}$, where $\overline{\mathbf{H}}_i={\left[\overline{\mathbf{h}}_{i,0}, \overline{\mathbf{h}}_{i,1}, \cdots, \overline{\mathbf{h}}_{i,N-1}\right]}^{\dag}\in\mathbb{C}^{N\times M},$ $\overline{\mathbf{h}}_{i,j}=\frac{\mathbf{h}_{i,j}}{\lVert{\mathbf{h}_{i,j}}\rVert},j\in{\left\{0,1,\cdots,N-1\right\}}$. 
The column vectors of $\mathbf{G}_i\in\mathbb{C}^{M\times {\left(M-N\right)}}$ constitute an orthogonal basis of the null space of $\overline{\mathbf{H}}_i^{\dag}$, i.e., $\mathbf{h}_{i,j}^{\dag}\mathbf{G}_i=\mathbf{0}$.

\subsection{Wyner's Wiretap Code and Secrecy Throughput}
\label{WynerCode}
We assume UAVs adopt the Wyner's wiretap encoding scheme. The transmission rate and redundant rate to secure the secret information are denoted by $\mathcal{R}_t$ and $\mathcal{R}_e,$ respectively, so the secrecy rate is $\mathcal{R}_s=\mathcal{R}_t-\mathcal{R}_e$. The probability that the capacity $\mathcal{C}_t$ of a legitimate channel is above the transmission rate $\mathcal{R}_t$ is the \emph{coverage probability} defined as 
\begin{equation}
\label{Pc_def}
\mathcal{P}_{c}\triangleq\mathbb{P}\left\{\mathcal{C}_t\geq\mathcal{R}_t\right\}\triangleq\mathbb{P}\left\{SINR\geq\beta_t\right\},
\end{equation}
where $\beta_t=2^{\mathcal{R}_t}-1$. The probability that the capacities of all wiretap channels are below the redundant rate $\mathcal{R}_e$ is the \emph{secrecy probability} defined as
\begin{equation}
\label{Ps_def} \mathcal{P}_{s}\triangleq\mathbb{E}_{\Phi_E}\left[\prod_{e\in\Phi_E}\mathbb{P}\left\{\mathcal{C}_{e}<\mathcal{R}_e\right\}\right].
\end{equation}

The secrecy throughput can be used to evaluate the secure transmission efficiency of the network. If a user can decode messages correctly and Eves can not decode secret messages to this user, the corresponding legitimate link is called a \emph{secrecy link}. The ST is defined as the average achievable secrecy rate by all the secrecy links in the unit area. For a network with the density of links being $\lambda,$ the ST is expressed as
\begin{equation}
\label{ST_def}
\mathcal{ST}\triangleq \lambda\mathcal{R}_s\mathcal{P}_{c}\mathcal{P}_{s}.
\end{equation}

In the following section, we provide an introduction to the MHCPP and elaborate how to use a non-homogeneous PPP to approximate it.

\section{Mat$\acute{\text{e}}$rn Hard-Core Point Process}\label{MHC}
The MHCPP is a repulsive point process, where the distance between each two nodes is no smaller than a predefined constant $d$, therefore it can characterize the minimum safety distance between two UAVs \cite{StoGeoWirlesNetVol1Theo}.  
The process of generating an MHCPP is: i) generate a PPP $\Phi_P$ with the intensity $\lambda_p$, ii) mark each point $u_i\in\Phi_P$ with an i.i.d. random variable $m_i$ uniformly distributed in $\left[0,1\right],$ iii) retain the point $u_i$ having the minimal mark compared to all the points in a circle centered at $u_i$ with the radius $d$. All the retained points form the MHCPP $\Phi_U$ with the intensity \cite{StoGeoWirlesNetVol1Theo}
\begin{equation}
\label{lambda_u_eq}
\lambda_u=\frac{1-e^{-\overline{K}}}{\pi d^2},
\end{equation}
where $\overline{K}=\lambda_p\pi d^2$. 

From the generation of $\Phi_U$, we find the locations of nodes are dependent. To derive the CP and SP of the typical user, we need to calculate the Laplace transform (LT) of the interference from UAVs in $\Phi_U$. However, the dependence of nodes' locations makes it hard to calculate the LT \cite{StoGeoWirlesNetVol2App}. According to Section 18.5 in \cite{StoGeoWirlesNetVol2App}, the MHCPP can be approximated by a non-homogeneous PPP with intensity $\lambda_p\mathcal{P}_r\left(r\right)$, where the probability $\mathcal{P}_r\left(r\right)=\mathbb{P}\left\{u_i\in\Phi_U\middle\vert u_0\in\Phi_U,u_i\in\Phi_P,{\left\lVert u_0-u_i\right\rVert}=r\right\}$. With this approximation, it is tractable to calculate the LT, and from Section \ref{Results}, the approximate results can well match with the simulation results. Therefore, we resort to approximate the MHCPP by a PPP. $\mathcal{P}_r\left(r\right)$ is derived in the following lemma.
\begin{lemma}
\label{Pr}
The probability that a point $u_i$ belongs to an MHCPP $\Phi_U$ when another point $u_0$ belongs to $\Phi_U$ and the distance between them is $r$ is given by
\begin{equation}
\label{PrEq}
\mathcal{P}_r\left(r\right)=
\left\{
\begin{array}{cl}
0, & 0<r<d \\
\frac{2}{\lambda_pV{\left(r\right)}-\overline{K}}{\left[1-\frac{\overline{K}{\left(1-e^{-\lambda_pV{\left(r\right)}}\right)}}{\lambda_pV{\left(r\right)}{\left(1-e^{-\overline{K}}\right)}}\right]}, & d\leq r<2d \\
\frac{1-e^{-\overline{K}}}{\overline{K}}, & r\geq 2d,
\end{array}
\right.
\end{equation}
where \cite{SecRateAnaUAVmmWavNetUsinMHCPP}
\begin{equation}
V\left(r\right)=
\left\{
\begin{array}{cl}
0, & 0<r<d \\
2\pi d^2-2d^2\cos^{-1}{\left(\frac{r}{2d}\right)}+r\sqrt{d^2-\frac{r^2}{4}}, & d\leq r<2d \\
2\pi d^2, & r\geq 2d.
\end{array}
\right.
\end{equation} 
\begin{IEEEproof}
From the definition of $\mathcal{P}_r\left(r\right)$, we know
\begin{equation}
\mathcal{P}_r\left(r\right)=\frac{\mathbb{P}{\left\{u_0\in\Phi_U,u_i\in\Phi_U\middle\vert u_i\in\Phi_P,{\left\lVert u_0-u_i\right\rVert}=r\right\}}}{\mathbb{P}{\left\{u_0\in\Phi_U\middle\vert u_i\in\Phi_P,{\left\lVert u_0-u_i\right\rVert}=r\right\}}}. \notag
\end{equation}
Denote marks of $u_0$ and $u_i$ by $m_0$ and $m_i$, respectively. By invoking Proposition 18.4.1. in \cite{StoGeoWirlesNetVol2App}, we can obtain 
\begin{align}
&\mathbb{P}{\left\{u_0\in\Phi_U,u_i\in\Phi_U\middle\vert m_0=t_0,m_i=t_i,u_i\in\Phi_P,{\left\lVert u_0-u_i\right\rVert}=r\right\}}= \notag \\
&\left\{
\begin{array}{cl}
0, & 0<r<d \\
\exp{\left(-\lambda_pt_0{\left(V{\left(r\right)}-\pi d^2\right)}-\lambda_pt_i\pi d^2\right)}, & d\leq r<2d \\
\exp{\left(-\lambda_pt_0\pi d^2-\lambda_pt_i\pi d^2\right)}, & r\geq2d,
\end{array}
\right. \notag
\end{align}
where the expectation with respect to (w.r.t.) $t_0$ and $t_i$ is
\begin{align}
\label{eq1}
&\mathbb{P}{\left\{u_0\in\Phi_U,u_i\in\Phi_U\middle\vert u_i\in\Phi_P,{\left\lVert u_0-u_i\right\rVert}=r\right\}}= \notag \\
&\left\{
\begin{array}{cl}
0, & 0<r<d \\
\frac{2}{\lambda_pV{\left(r\right)}-\overline{K}}{\left[\frac{1-e^{-\overline{K}}}{\overline{K}}-\frac{1-e^{-\lambda_pV{\left(r\right)}}}{\lambda_pV{\left(r\right)}}\right]}, & d\leq r<2d \\
\frac{\left(1-e^{-\overline{K}}\right)^2}{\overline{K}^2}, & r\geq2d.
\end{array}
\right. 
\end{align}
The probability that $u_0\in\Phi_U$ when $m_0=t_0$ is
\begin{equation}
\mathbb{P}\left\{u_0\in\Phi_U\middle\vert m_0=t_0\right\}=\sum\nolimits_{n=0}^{\infty}\frac{\overline{K}^ne^{-\overline{K}}}{n!}\cdot{\left(1-t_0\right)}^n=e^{-t_0\overline{K}}. \notag
\end{equation}
Therefore, we have
\begin{align}
\label{eq2}
&\quad\mathbb{P}{\left\{u_0\in\Phi_U\middle\vert u_i\in\Phi_P,{\left\lVert u_0-u_i\right\rVert}=r\right\}} \notag \\
&=
\left\{
\begin{array}{cl}
\int_0^1e^{-t_0\overline{K}}{\left(1-t_0\right)}dt_0, & 0<r<d \\
\int_0^1e^{-t_0\overline{K}}dt_0, & r\geq d
\end{array}
\right. \notag \\
&=
\left\{
\begin{array}{cl}
\frac{\overline{K}+e^{-\overline{K}}-1}{\overline{K}^2}, & 0<r<d \\
\frac{1-e^{-\overline{K}}}{\overline{K}}, & r\geq d.
\end{array}
\right.
\end{align}
\eqref{eq1} divided by \eqref{eq2} completes the proof.
\end{IEEEproof}
\end{lemma}
With $\mathcal{P}_r{\left(r\right)}$,  
the MHCPP $\Phi_U$ can be approximated by a PPP with the intensity $\lambda_p\mathcal{P}_r{\left(r\right)}$.

\section{Coverage Probability}\label{CovPro}
In this section, we analyze the CP of users. Without loss of generality, we perform analysis for a randomly chosen \emph{typical user} $x_{u_0,0}$ served by the UAV $u_0$.

From \eqref{signal}, the signal-to-interference-and-noise ratio (SINR) of the typical user is
\begin{equation}
SINR_x=\frac{P_shL_p{\left(l_{0,0}\right)}}{I_{x,L}+I_{x,N}+\sigma_x^2},
\end{equation} 
where $h={\left\lVert\mathbf{h}_{0,0}^{\dag}\mathbf{W}_0\right\rVert}^2$, $I_{x,L}=\sum_{u_i\in\Phi_U^L\backslash u_0}P_iL_{p,L}{\left(l_{i,0}\right)}$ and $I_{x,N}=\sum_{u_i\in\Phi_U^N\backslash u_0}P_iL_{p,N}{\left(l_{i,0}\right)}$ are the interferences from UAVs with LoS and NLoS links, respectively,  $P_i=P_sg_{I,i}+P_ng_{N,i}$, $P_s=\frac{\phi P}{N}$, $P_n=\frac{{\left(1-\phi\right)}P}{M-N}$, $g_{I,i}={\left\Vert\mathbf{g}_{i,0}^{\dag}\mathbf{W}_i\right\rVert}^2$, $g_{N,i}={\left\lVert\mathbf{g}_{i,0}^{\dag}\mathbf{G}_i\right\rVert}^2$, $l_{0,0}$ and $l_{i,0}$ are the horizontal distances from the serving UAV $u_0$ and the interfering UAV $u_i$ to the typical user $x_{u_0,0}$, respectively.
According to the definition of CP given in \eqref{Pc_def}, the CP of the typical user $x_{u_0,0}$ is derived in the following theorem.
\begin{theorem}
\label{CP}
The CP of a typical user is given by
\begin{equation}
\label{Pc}
\mathcal{P}_c=\int_0^{\infty}{\left[\mathcal{P}_{c,L}{\left(l\right)}\mathcal{P}_L{\left(l\right)}+\mathcal{P}_{c,N}{\left(l\right)}\mathcal{P}_N{\left(l\right)}\right]}f_{l_{0,0}}{\left(l\right)}dl,
\end{equation}
where 
\begin{align}
\mathcal{P}_{c,Q}&=
e^{-s_Q\sigma_x^2}\sum_{k=0}^{M-N}\frac{s_Q^k}{k!}\sum_{m=0}^k\binom{k}{m}{\left(-1\right)}^m\sigma_x^{2\left(k-m\right)}\sum_{n=0}^m\binom{m}{n} \notag\\
&\times\mathcal{L}_{I_{x,L}}^{\left(n\right)}{\left(s_Q\right)}\mathcal{L}_{I_{x,N}}^{\left(m-n\right)}{\left(s_Q\right)}, \notag
\end{align}
$Q\in{\left\{L,N\right\}}$ is the CP conditioned on the LoS or NLoS legitimate link, $s_Q=\frac{\beta_t}{P_s\eta_Q\xi R{\left(l\right)}^{-\alpha_Q}}$,
$\mathcal{L}_{I_{x,L}}{\left(s_Q\right)}$ and $\mathcal{L}_{I_{x,N}}{\left(s_Q\right)}$ are the LTs of $I_{x,L}$ and $I_{x,N}$, respectively, $\mathcal{L}_{I_{x,L}}^{\left(q\right)}{\left(s_Q\right)}$ and $\mathcal{L}_{I_{x,N}}^{\left(q\right)}{\left(s_Q\right)}$ are the $q$-order derivatives of $\mathcal{L}_{I_{x,L}}{\left(s_Q\right)}$ and $\mathcal{L}_{I_{x,N}}{\left(s_Q\right)}$ w.r.t. $s_Q$, respectively, $\mathcal{P}_N{\left(l\right)}=1-\mathcal{P}_L{\left(l\right)}$ is the probability of NLoS links, and $f_{l_{0,0}}{\left(l\right)}=\frac{l}{\sigma^2}\exp{\left(-\frac{l^2}{2\sigma^2}\right)}$.
\begin{IEEEproof}
Please see Appendix \ref{Prf_CP}.
\end{IEEEproof}
\end{theorem}
Here the LT of the interference $I$ is defined as $\mathcal{L}_{I}{\left(s\right)}=\mathbb{E}{\left[e^{-sI}\right]}$. $\mathcal{L}_{I_{x,L}}{\left(s_Q\right)}$ and $\mathcal{L}_{I_{x,N}}{\left(s_Q\right)}$ are given in the following lemma.
\begin{lemma}
The LTs of the interference $I_{x,L}$ and $I_{x,N}$ at a typical user are given by
\begin{align}
&\mathcal{L}_{I_{x,Q}}{\left(s\right)} \notag \\
&=\exp{\left[-\lambda_p\int_{0}^{2\pi}\int_{d}^{\infty}\mathcal{P}_Q{\left(l_{i,0}\right)}\mathcal{P}_r{\left(r_{0,i}\right)}{\left(1-\Theta_Q{\left(r_{0,i},\beta\right)}\right)}r_{0,i}dr_{0,i}d\beta\right]} \notag \\
&=\exp{\left[\Omega_Q{\left(s\right)}\right]},
\end{align}
where $Q\in{\left\{L,N\right\}}$, $\Theta_Q{\left(r_{0,i},\beta\right)}=$
\begin{align}
&\left\{
\begin{array}{cl}
\sum\limits_{n=0}^{M-N-1}\frac{C_{N+n-1}^nP_n^{2N-M}}{{\left(-\zeta\right)}^n{\left(P_n-P_s\right)}^N}{\left[\frac{1}{\tau_{2,Q}^{M-N-n}}-\sum\limits_{m=0}^{N+n-1}\frac{C_{M-N-n-1+m}^m\zeta^m}{\tau_{1,Q}^{M-N-n+m}}\right]}, & 0<\phi<\frac{N}{M} \\
\sum\limits_{n=0}^{N-1}\frac{C_{M-N+n-1}^nP_s^{M-2N}}{\zeta^n{\left(P_s-P_n\right)}^{M-N}}{\left[\frac{1}{\tau_{1,Q}^{N-n}}-\sum\limits_{m=0}^{M-N+n-1}\frac{C_{N-n-1+m}^m{\left(-\zeta\right)}^m}{\tau_{2,Q}^{N-n+m}}\right]}, & \frac{N}{M}<\phi<1 \\
\frac{1}{{\left(P_s\tau_{1,Q}\right)}^M}, & \phi=\frac{N}{M}, \notag
\end{array}
\right.
\end{align}
$\tau_{1,Q}=P_s^{-1}+s\eta_Q\xi R{\left(l_{i,0}\right)}^{-\alpha_Q}$, $\tau_{2,Q}=P_n^{-1}+s\eta_Q\xi R{\left(l_{i,0}\right)}^{-\alpha_Q}$, $\zeta=\frac{P_n-P_s}{P_sP_n}$, $l_{i,0}=\sqrt{l_{0,0}^2+r_{0,i}^2-2l_{0,0}r_{0,i}\cos\beta}$, and $r_{0,i}$ is the distance from the UAV $u_i$ to the UAV $u_0$.
\begin{IEEEproof}
According to the definition and the independence in the small-scale fading of the interferences from different UAVs, $\mathcal{L}_{I_{x,Q}}{\left(s\right)}=\mathbb{E}_{\Phi_U^Q}{\left[\prod_{u_i\in\Phi_U^Q\backslash u_0}\mathbb{E}_{P_i}{\left[\exp{\left(-sP_iL_{p,Q}{\left(l_{i,0}\right)}\right)}\right]}\right]}$. Denote $\mathbb{E}_{P_i}{\left[\exp{\left(-sP_iL_{p,Q}{\left(l_{i,0}\right)}\right)}\right]}$ by $\Theta_Q{\left(r_{0,i},\beta\right)}$. As explained in Section \ref{MHC}, we use a PPP with the intensity $\lambda_p\mathcal{P}_r{\left(r_{0,i}\right)}$ to approximate $\Phi_U$. Considering the probabilities of LoS and NLoS links, $\Phi_U^Q$, $Q\in{\left\{L,N\right\}}$ can be approximated by a PPP with the intensity $\lambda_p\mathcal{P}_Q{\left(l_{i,0}\right)}\mathcal{P}_r{\left(r_{0,i}\right)}$. The result then can be derived using the probability generation functional (PGFL) of PPPs \cite{StoGeoWirlesNetVol2App}.

For $\Theta_Q{\left(r_{0,i},\beta\right)}$, with $g_{I,i}\sim\Gamma{\left(N,1\right)}$, $g_{N,i}\sim\Gamma{\left(M-N,1\right)}$ and the property of the Gamma distribution, we have $X_i=P_sg_{I,i}\sim\Gamma{\left(N,P_s\right)}$ and $Y_i=P_ng_{N,i}\sim\Gamma{\left(M-N,P_n\right)}$. Since $X_i$ and $Y_i$ are independent, with the aid of \cite[eq. (3.381.8.*)]{IntegralsTable}, we derive the PDF of $P_i=X_i+Y_i$ as
\begin{equation}
f_{P_i}{\left(p\right)}
=\left\{
\begin{array}{cl}
\frac{\sum_{n=0}^{M-N-1}C_{M-N-1}^n{\left(-1\right)}^np^{M-N-1-n}e^{-\frac{p}{P_n}}\gamma{\left(N+n,\zeta p\right)}}{P_s^NP_n^{M-N}\zeta^{N+n}\Gamma{\left(N\right)}\Gamma{\left(M-N\right)}}, & 0<\phi<\frac{N}{M} \\
\frac{\sum_{n=0}^{N-1}C_{N-1}^n{\left(-1\right)}^np^{N-1-n}e^{-\frac{p}{P_s}}\gamma{\left(M-N+n,-\zeta p\right)}}{P_s^NP_n^{M-N}{\left(-\zeta\right)}^{M-N+n}\Gamma{\left(N\right)}\Gamma{\left(M-N\right)}}, & \frac{N}{M}<\phi<1 \\
\frac{p^{M-1}e^{-\frac{p}{P_s}}}{P_s^M\Gamma{\left(M\right)}}, & \phi=\frac{N}{M}. \notag
\end{array}
\right.
\end{equation}
By \cite[eq. (3.381.4)]{IntegralsTable}, $\int_0^{\infty}\exp{\left(-spL_{p,Q}{\left(l_{i,0}\right)}\right)}f_{P_i}{\left(p\right)}dp$ is derived. This completes the proof.
\end{IEEEproof}
\end{lemma}

Then $\mathcal{L}_{I_{x,Q}}^{\left(q\right)}{\left(s\right)}$ is derived in the following lemma.
\begin{lemma}
The $q$-order derivative of $\mathcal{L}_{I_{x,Q}}{\left(s\right)}$ is given by
\begin{equation}
\mathcal{L}_{I_{x,Q}}^{\left(q\right)}{\left(s\right)}=\sum\nolimits_{p=0}^{q-1}\binom{q-1}{p}\mathcal{L}_{I_{x,Q}}^{\left(p\right)}{\left(s\right)}\Omega_Q^{\left(q-p\right)}{\left(s\right)},
\end{equation}
where $\Omega_Q^{\left(k\right)}{\left(s\right)}=\lambda_p\int_{0}^{2\pi}\int_d^{\infty}\mathcal{P}_Q{\left(l_{i,0}\right)}\mathcal{P}_r{\left(r_{0,i}\right)}\Xi_Q{\left(k,r_{0,i},\beta\right)}r_{0,i}dr_{0,i}d\beta$,
\begin{align}
&\Xi_Q{\left(k,r_{0,i},\beta\right)}= \notag \\
&\left\{
\begin{array}{cl}
\sum\limits_{n=0}^{M-N-1}\varrho_1^Q{\left[\varphi_2^Q{\left(M-N,k\right)}-\sum\limits_{m=0}^{N+n-1}\frac{\varphi_1^Q{\left(M-N,k+m\right)}\zeta^m}{m!}\right]}, & 0<\phi<\frac{N}{M} \\
\sum\limits_{n=0}^{N-1}\varrho_2^Q{\left[\varphi_1^Q{\left(N,k\right)}-\sum\limits_{m=0}^{M-N+n-1}\frac{\varphi_2^Q{\left(N,k+m\right)}{\left(-\zeta\right)}^m}{m!}\right]}, & \frac{N}{M}<\phi<1 \\
\frac{A_{M+k-1}^k{\left(-L_{p,Q}{\left(l_{i,0}\right)}\right)}^k}{P_s^M\tau_{1,Q}^{M+k}}, & \phi=\frac{N}{M}, \notag
\end{array}
\right.
\end{align}
$\varrho_1^Q=\frac{C_{N+n-1}^nP_n^{2N-M}{\left(-L_{p,Q}{\left(l_{i,0}\right)}\right)}^k}{{\left(-\zeta\right)}^n{\left(P_n-P_s\right)}^N}$, $\varrho_2^Q=\frac{C_{M-N+n-1}^nP_s^{M-2N}{\left(-L_{p,Q}{\left(l_{i,0}\right)}\right)}^k}{\zeta^n{\left(P_s-P_n\right)}^{M-N}}$, $\varphi_1^Q{\left(K,k\right)}=\frac{A_{K-n-1+k}^{k}}{\tau_{1,Q}^{K-n+k}}$, and $\varphi_2^Q{\left(K,k\right)}=\frac{A_{K-n-1+k}^{k}}{\tau_{2,Q}^{K-n+k}}$.
\end{lemma}

\section{Secrecy Probability and Secrecy Throughput}\label{SPST}
In this section, we first analyze the SP of users, and then we derive the ST of the network. Due to the randomness in the locations of Eves, we assume that Eves do not collude. To evaluate the lower bound of the secrecy performance, Eves are assumed to have the ability of multi-user decoding, e.g., by using the technology of serial interference cancellation. Therefore, Eves are only interfered with the AN and the thermal noise. The SINR of an Eve $e\in\Phi_E$ is
\begin{equation}
SINR_e=\frac{P_s\lvert{\mathbf{g}_{0,e}^{\dag}\mathbf{w}_{0,0}}\rvert^2L_p{\left(l_{0,e}\right)}}{I_{e,L}+I_{e,N}+\sigma_e^2},
\end{equation}
where $I_{e,L}=\sum_{u_i\in\Phi_U^L}P_n\lVert{\mathbf{g}_{i,e}^{\dag}\mathbf{G}_i}\rVert^2L_{p,L}{\left(l_{i,e}\right)}$ and $I_{e,N}=\sum_{u_i\in\Phi_U^N}P_n\lVert{\mathbf{g}_{i,e}^{\dag}\mathbf{G}_i}\rVert^2L_{p,N}{\left(l_{i,e}\right)}$ are interferences from LoS and NLoS links, respectively. The SP is given as follows.
\begin{theorem}
The SP of a typical user is given by
\begin{equation}
\label{Ps}
\mathcal{P}_s=\exp{\left[-\frac{2\pi\lambda_e}{{\left(1+\frac{\beta_eP_n}{P_s}\right)}^{M-N}}\int\nolimits_0^{\infty}\sum_{Q\in{\left\{L,N\right\}}}\mu_Q{\left(l_{0,e}\right)}\Psi_Q{\left(l_{0,e}\right)}l_{0,e}dl_{0,e}\right]},
\end{equation}
where $\mu_Q{\left(l_{0,e}\right)}=\exp{\left(-\frac{\beta_e\sigma_e^2}{P_sL_{p,Q}{\left(l_{0,e}\right)}}\right)}\mathcal{P}_Q{\left(l_{0,e}\right)}$, $\Psi_Q{\left(l_{0,e}\right)}=\exp{\left(-\lambda_p\int\limits_0^{2\pi}\int\limits_d^{\infty}\mathcal{P}_r{\left(r_{0,i}\right)}\varepsilon_Q{\left(l_{0,e},r_{0,i},\beta\right)}r_{0,i}dr_{0,i}d\beta\right)}$, $\varepsilon_Q{\left(l_{0,e},r_{0,i},\beta\right)}=1-\frac{\mathcal{P}_L{\left(l_{i,e}\right)}}{{\left(1+\frac{\beta_eP_nL_{p,L}{\left(l_{i,e}\right)}}{P_sL_{p,Q}{\left(l_{0,e}\right)}}\right)}^{M-N}}-\frac{\mathcal{P}_N{\left(l_{i,e}\right)}}{{\left(1+\frac{\beta_eP_nL_{p,N}{\left(l_{i,e}\right)}}{P_sL_{p,Q}{\left(l_{0,e}\right)}}\right)}^{M-N}}$, $\beta_e=2^{\mathcal{R}_e}-1$, and the horizontal distance from the $i$-th UAV to an Eve is $l_{i,e}=\sqrt{l_{0,e}^2+r_{0,i}^2-2l_{0,e}r_{0,i}\cos\beta}$.
\begin{IEEEproof}
Knowing $\lvert{\mathbf{g}_{0,e}^{\dag}\mathbf{w}_{0,0}}\rvert^2\sim\exp{\left(1\right)}$ and $\lVert{\mathbf{g}_{i,e}^{\dag}\mathbf{G}_i}\rVert^2\sim\Gamma{\left(M-N,1\right)}$, using the PPP with the density $\lambda_p\mathcal{P}_r{\left(r_{0,i}\right)}$ to approximate the MHCPP $\Phi_U$ and  invoking the PGFL of the PPP, we can derive the analytical result.
\end{IEEEproof}
\end{theorem}

With the CP and SP, the ST is given as follows.
\begin{theorem}
The ST of the network is given by
\begin{equation}
\label{STeq}
\mathcal{ST}=\frac{\lambda_pN\mathcal{P}_c\mathcal{P}_s\mathcal{R}_s{\left(1-e^{-\overline{K}}\right)}}{\overline{K}}.
\end{equation}
\begin{IEEEproof}
According to the definition of ST, we can calculate the average achievable secrecy rate by all the secrecy links in a circle with the radius $\rho$ and then divide the result by the circle area to derive ST. However, as stated in Lemma \ref{Pr}, the probability that a point belongs to $\Phi_U$ depends on the distance between this point and another point belonging to $\Phi_U$. Thus we let $\rho\rightarrow\infty$ and the specific proof is as follows:
\begin{align}
\mathcal{ST} &= \lim_{\rho\rightarrow\infty}\frac{1}{\pi\rho^2}\mathbb{E}
\left[\sum_{u_i\in\Phi_U\cap\mathcal{B}{\left(u_0,\rho\right)}}\sum\nolimits_{j=0}^{N-1}\mathbb{I}{\left(SINR_{x_{u_i,j}}\geq\beta_t\right)}\right. \notag \\
&\qquad\qquad\qquad\qquad\qquad\qquad\left.\times\mathbb{I}{\left(\bigcap\limits_{e\in\Phi_E}SINR_{e,x_{u_i,j}}<\beta_e\right)}\mathcal{R}_s\right] \notag \\
&=\lim_{\rho\rightarrow\infty}\frac{1}{\pi\rho^2}\mathbb{E}{\left[\sum_{u_i\in\Phi_U\backslash u_0\cap\mathcal{B}{\left(u_0,\rho\right)}}N\mathcal{P}_c\mathcal{P}_s\mathcal{R}_s+N\mathcal{P}_c\mathcal{P}_s\mathcal{R}_s\right]} \notag \\
&=\lim_{\rho\rightarrow\infty}\frac{1}{\pi\rho^2}{\left(\int_d^{\rho}\int_0^{2\pi}\lambda_p\mathcal{P}_r{\left(r\right)}N\mathcal{P}_c\mathcal{P}_s\mathcal{R}_srdrd\beta+N\mathcal{P}_c\mathcal{P}_s\mathcal{R}_s\right)} \notag \\
&=\frac{\lambda_pN\mathcal{P}_c\mathcal{P}_s\mathcal{R}_s{\left(1-e^{-\overline{K}}\right)}}{\overline{K}}. \notag
\end{align} 
This completes the proof.
\end{IEEEproof}
\end{theorem}

\begin{figure}[!t]
\begin{center}
\centering
\includegraphics[width=3.1 in]{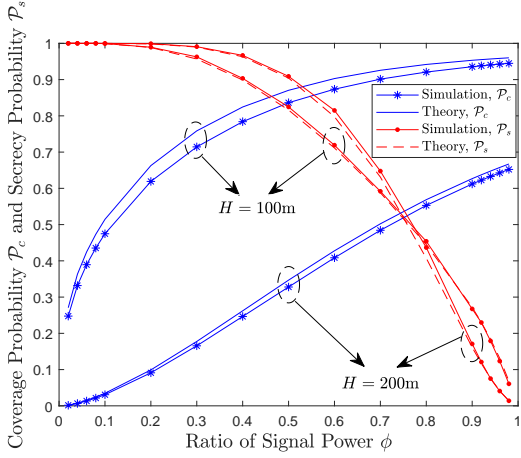}
\end{center}
\caption{$\mathcal{P}_c$ in \eqref{Pc} and $\mathcal{P}_s$ in \eqref{Ps} vs. $\phi$ with $d=50$m, $\lambda_u=8/10^6$m$^{-2}$, $\sigma=20$m, $M=8$, and $N=4$.}
\label{PcPs_H_phi}
\vspace{-6.5mm}
\end{figure}

\section{Results and Discussions}\label{Results}
In this section, we show some numerical results, which validate our theoretical derivations. In the simulations, we set $\alpha_L=2.5$, $\alpha_N=2.8$, $\eta_L=-1.6$dB, $\eta_N=-23$dB, $\xi=-40$dB, $a=11.95$, $b=0.136$, $\lambda_e=8/10^6$m$^{-2}$, $P=5$W, $\sigma_x^2=\sigma_e^2=-100$dBm, $\mathcal{R}_t=0.8$bits/s/Hz and $\mathcal{R}_e=0.5\mathcal{R}_t$.

Fig.~\ref{PcPs_H_phi} depicts the CP and SP of the typical user versus the ratio $\phi$ of the information-bearing signal power to the total transmission power of UAVs for various heights $H$ of UAVs. It is observed that the theoretical results can approximate the simulation results well. With $\phi$ increasing, the CP increases while the SP decreases, for more power used for information-bearing signals and less power used for AN. For the CP, the theoretical results approximate the simulation results better when $H$ becomes larger. When $\phi$ is relatively small, sufficient power is used for AN. Therefore, increasing $H$ has more impact on decreasing the signal power than decreasing the AN power received by Eves, leading to the SP increasing. When $\phi$ is relatively large, little power is used for AN. Increasing $H$ will further decrease the AN power received by Eves, so the SP decreases with $H$.

\begin{figure}[!t]
\begin{center}
\centering
\includegraphics[width=3.1 in]{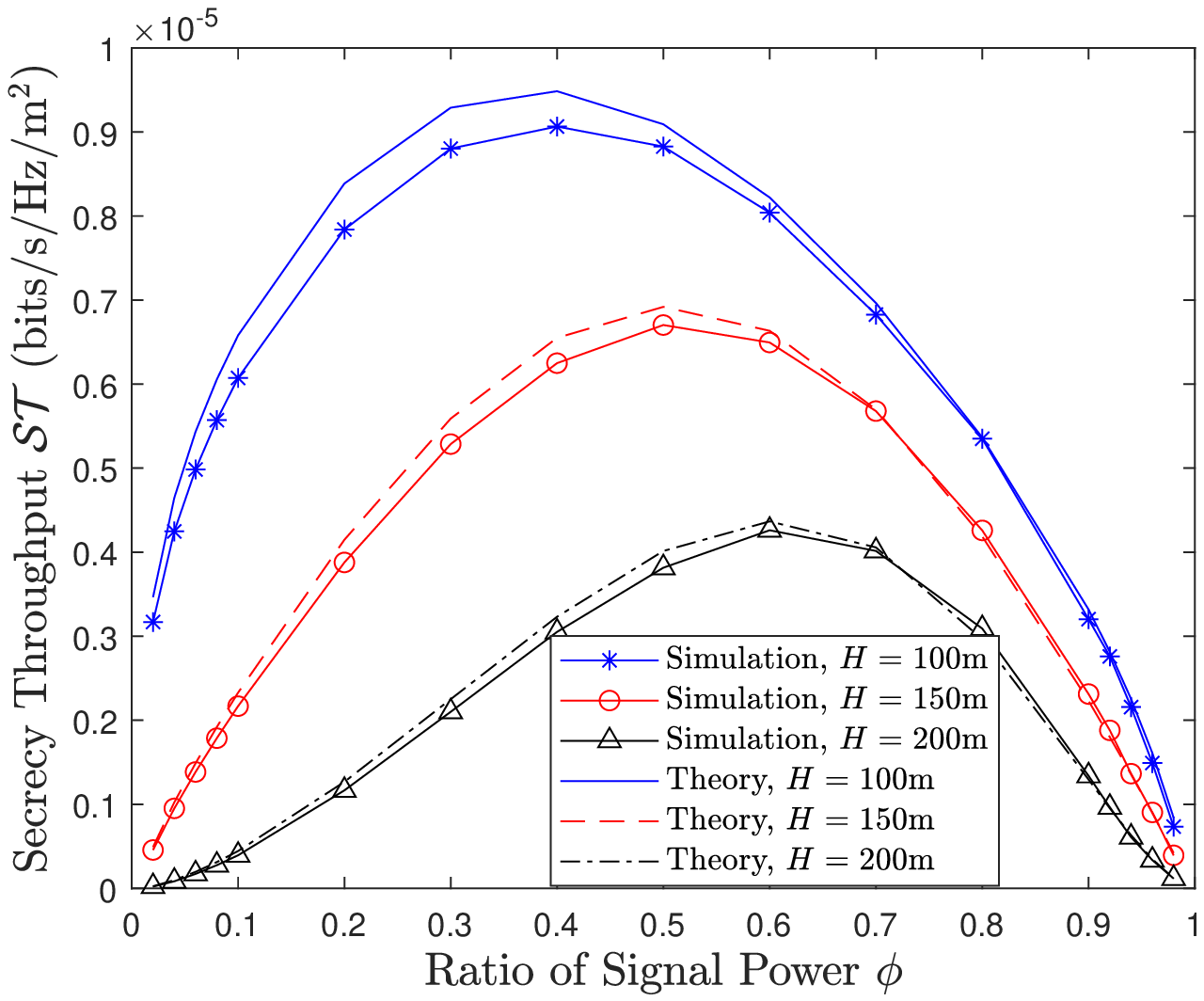}
\end{center}
\caption{$\mathcal{ST}$ in \eqref{STeq} vs. $\phi$ with $d=50$m, $\lambda_u=8/10^6$m$^{-2}$, $\sigma=20$m, $M=8$, and $N=4$.}
\label{ST_H_phi}
\vspace{-1mm}
\end{figure}

Fig.~\ref{ST_H_phi} plots the ST of the network versus $\phi$ for various $H$. The theoretical results can approximate the simulation results well. We can see the ST first increases and then decreases with $\phi$ increasing, which is the balance of more AN power to combat Eves and more signal power to cover users. Besides, the ST decreases and the optimal $\phi$ maximizing the ST increases with $H$ increasing. This indicates when UAVs are deployed at a higher altitude, more power should be allocated for transmitting signals to maximize the ST of the network.

\begin{figure}[!t]
\begin{center}
\centering
\includegraphics[width=3.1 in]{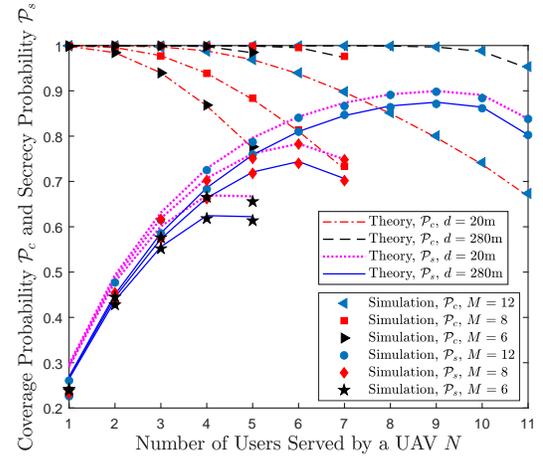}
\end{center}
\caption{$\mathcal{P}_c$ in \eqref{Pc} and $\mathcal{P}_s$ in \eqref{Ps} vs. $N$ with $H=100$m, $\lambda_u=4/10^6$m$^{-2}$, $\sigma=10$m, and $\phi=0.5$.}
\label{PcPs_d_M_N}
\vspace{-6.5mm}
\end{figure}

Fig.~\ref{PcPs_d_M_N} depicts the CP and SP of the typical user versus the number $N$ of users served by each UAV with various minimum safety distances $d$ and the numbers of antennas $M$ at UAVs. We can observe the accuracy of theoretical results are further validated in various scenarios. When $d$ decreases, the CP decreases but the SP increases, because UAVs are allowed to be closer when $d$ is smaller, leading to shorter interference link distances at the user and Eves. If $N$ increases, the CP will decrease since the signal power at each user gets lower. Likewise, the signal power at an Eve also decreases, so the SP first increases with $N$. Meanwhile, the AN power at an Eve has a tendency to decrease. When $N$ gets very large, the impact of lower AN power exceeds that of lower signal power, resulting in the decreased SP. With $M$ increasing, users receive more signal power, so the CP increases. There are also more degrees of freedom to combat Eves with more antennas at UAVs, thus the SP increases. When $N$ is relatively small, Eves can receive enough signal power, so the improvement of SP with larger $M$ is not so obvious. In other words, the advantage of more antennas is more visible when each UAV serves more users.

\begin{figure}[!t]
\begin{center}
\centering
\includegraphics[width=3.1 in]{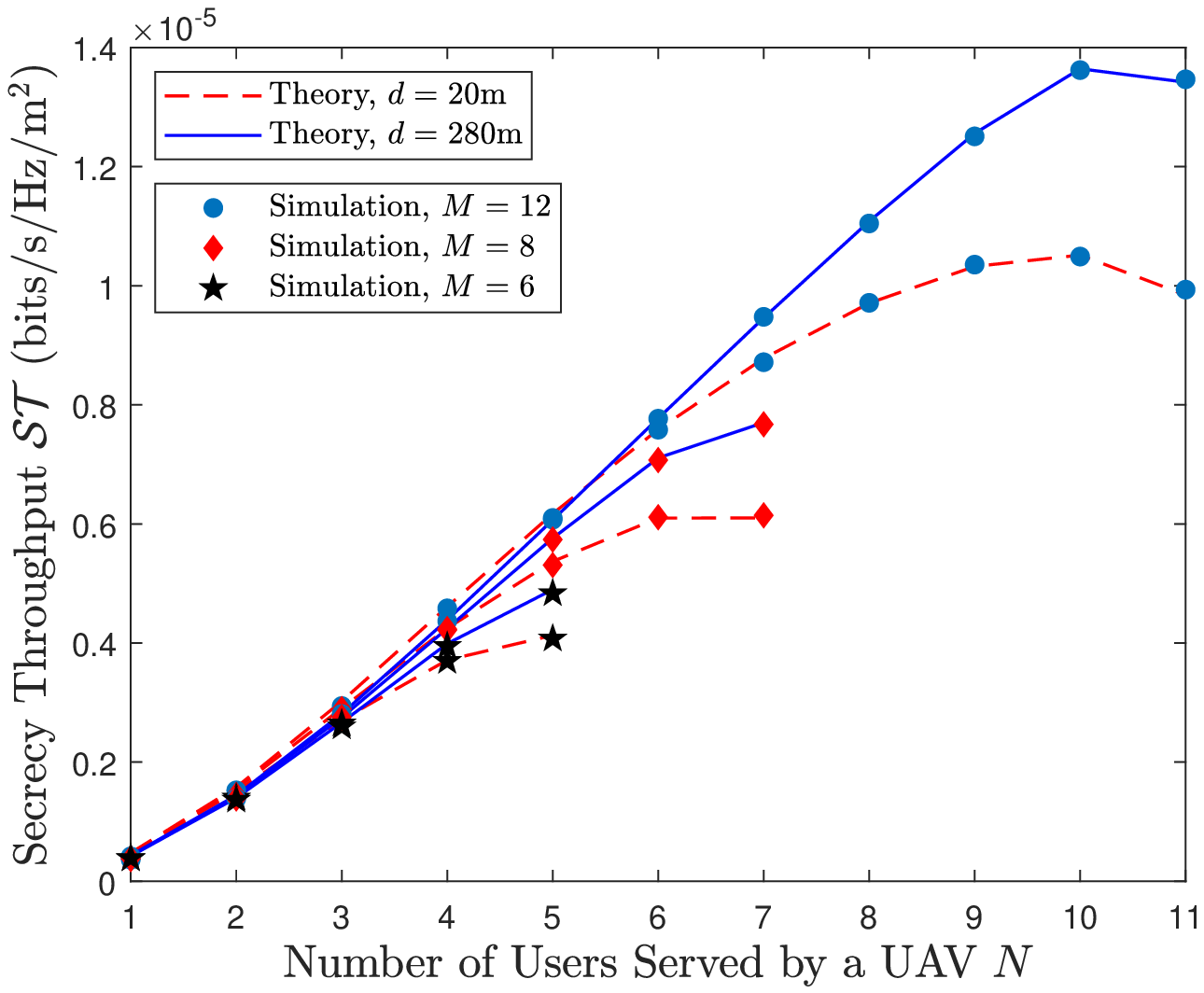}
\end{center}
\caption{$\mathcal{ST}$ in \eqref{STeq} vs. $N$ with $H=100$m, $\lambda_u=4/10^6$m$^{-2}$, $\sigma=10$m, and $\phi=0.5$.}
\label{ST_d_M_N}
\vspace{-6.5mm}
\end{figure}

Fig.~\ref{ST_d_M_N} plots the ST of the network versus $N$ for different $d$ and $M$. It also shows the accuracy of the approximation. We can see the ST increases with $d$, $M$ and $N$. This indicates that to achieve a larger ST, we can enlarge the minimum safety distance between two UAVs under some SP constraint and equip UAVs with more antennas to serve more users. When each UAV serves few users, the ST improves little with $d$ and $M$. In this situation, UAVs can have smaller minimum safety distances to improve SP and possess fewer antennas to save the computational overhead.

\section{Conclusions}\label{Conclusions}
In this correspondence, we analyzed the coverage and secrecy performances of a UAV network, where the locations of UAVs and ground users are modeled as an MHCPP and a PCP, respectively, and UAVs adopt ZF precoding to serve multiple users and AN-aided transmission strategy to combat Eves. We derived the approximations for the CP and SP of the typical user and the ST of the network. The simulation results can be well approximated by the theoretical results. The main conclusions are: (i) the SP will increase with $H$ when enough power is used for AN, while the SP will decrease with $H$ when most power is used for signals; (ii) the ST first increases and then decreases with $\phi$, while the ST decreases and the optimal $\phi$ maximizing the ST increases with $H$; (iii) the CP increases with $d$ and decreases with $N$, while the SP decreases with $d$ and first increases and then decreases with $N$; (iv) the ST increases with $d$, $M$ and $N$.

\appendices

\section{Proof of Theorem \ref{CP}}
\label{Prf_CP}
The main link is possible to be a LoS or NLoS link, therefore
$\mathcal{P}_c=\int_0^{\infty}{\left[\mathcal{P}_{c,L}{\left(l\right)}\mathcal{P}_L{\left(l\right)}+\mathcal{P}_{c,N}{\left(l\right)}\mathcal{P}_N{\left(l\right)}\right]}f_{l_{0,0}}{\left(l\right)}dl.$ 
Conditioned on the main link being a LoS link, $\mathcal{P}_{c,L}{\left(l\right)}$ is 
\begin{align}
\mathcal{P}_{c,L}{\left(l\right)} &= \mathbb{E}{\left[\mathbb{P}{\left\{h\geq\frac{{\left(I_{x,L}+I_{x,N}+\sigma_x^2\right)}\beta_t}{P_s\eta_L\xi R{\left(l\right)}^{-\alpha_L}}\right\}}\right]} \notag \\
&\overset{\text{(a)}}{=} \mathbb{E}{\left[e^{-s_L{\left(I_{x,L}+I_{x,N}+\sigma_x^2\right)}}\sum_{k=0}^{M-N}\frac{s_L^k{\left(I_{x,L}+I_{x,N}+\sigma_x^2\right)}^k}{k!}\right]} \notag \\
&\overset{\text{(b)}}{=} e^{-s_L\sigma_x^2}\sum_{k=0}^{M-N}\frac{s_L^k}{k!}\sum_{m=0}^k\binom{k}{m}{\left(-1\right)}^m\sigma_x^{2\left(k-m\right)}\sum_{n=0}^m\binom{m}{n} \notag \\
&\times\mathbb{E}{\left[{\left(-I_{x,L}\right)}^ne^{-s_LI_{x,L}}\right]}\mathbb{E}{\left[{\left(-I_{x,N}\right)}^{m-n}e^{-s_LI_{x,N}}\right]}, \notag
\end{align}
where (a) follows from $h\sim\Gamma{\left(M-N+1,1\right)}$ and using \cite[eq. (3.351.3)]{IntegralsTable} to calculate the integral. By using a PPP to approximate the MHCPP $\Phi_U$, $I_{x,L}$ and $I_{x,N}$ are independent, which leads to (b). Due to $\mathcal{L}_I^{\left(m\right)}{\left(s\right)}=\mathbb{E}{\left[{\left(-I\right)}^me^{-sI}\right]}$, we can get the result. The derivation of $\mathcal{P}_{c,N}{\left(l\right)}$ is similar.

\end{document}